# Polyhydrogenated Graphene – Excited State Dynamics in Photo- and Electroactive 2D-Domains

Volker Strauss[a], Ricarda A. Schäfer[b], Frank Hauke[b], Andreas Hirsch[b*], Dirk M. Guldi[a*]

[a] Department of Chemistry and Pharmacy & Interdisciplinary Center for Molecular Materials (ICMM), Friedrich-Alexander University of Erlangen-Nürnberg, Egerlandstr. 3, 91058 Erlangen, Germany

[b] Department of Chemistry and Pharmacy and Joint Institute of Advanced Materials and Processes (ZMP), Friedrich-Alexander University of Erlangen-Nürnberg, Henkestr. 42, 91054 Erlangen (Germany)

**ABSTRACT:** Understanding the phenomenon of intense photoluminescence in carbon materials such as hydrogenated graphene, graphene nanoribbons, etc. is at the forefront of investigations. In this study, six different types of hydrogenated graphene (phG) produced from different starting materials, were fully characterized in terms of structure and spectroscopy. Comprehensive photoluminescence lifetime analyses of phGs were conducted by combining time-correlated single-photon counting spectroscopy with steady-state fluorescence spectroscopy and femtosecond transient absorption spectroscopy. The conclusion drawn from these assays is that graphene islands with diameters in the range from 1.1 to 1.75 nm reveal band gap photoluminescence between 450 and 800 nm. As a complement, phGs were implemented in hybrids with water-soluble electron accepting perylenediimides (PDI). By virtue of mutual π-stacking and charge transfer interactions with graphene islands, PDIs assisted in stabilizing aqueous dispersion of phG. Implicit in these ground state interactions is the formation of 300 ps lived charge separated states once photoexcited.

Introduction

Hydrogenated graphene or graphane materials, that is, fully hydrogenated graphene, have drawn considerable attention in the area of hydrogen storage.[1-6] In addition, a large number of studies highlight the outstanding electronic properties of hydrogenated graphene etc.[5,7-10] During the course of graphene hydrogenation a sizeable bandgap develops which is intrinsically a zero bandgap material.[5,9,11-15] Such a bandgap opening has first been theoretically predicted and then been experimentally confirmed. This happens regardless of the degree or mechanism of hydrogenation.[11] For graphane as a two-dimensional semiconductor a bandgap of at least 3.5 eV has been predicted.[1,4,7] Still, a distinct photoluminescence is seen in the visible and the near infrared regions of the solar spectrum.[16-18] As a matter of fact, photoluminescence is also detectable in the solid state.

Lately, a general and controversial discussion emerged about the origin of photoluminescence in the different carbon materials. For example, a variety of nano-graphenic emitters, so called graphene quantum dots or carbon nanodots, have been fabricated with rather strong photoluminescence in the visible and the near infrared regions of the solar spectrum.[19-21] It is commonly accepted that the photoluminescence in these nano-graphenic emitters is trap based.[22] Notably, excitonic bandgap emission due to quantum confinement leads to the characteristic near infrared photoluminescence of single-walled carbon nanotubes (SWCNT).[23,24] Tailoring the defects along SWCNTs enables the selective tuning of the photoluminescent properties.[25,26] Interestingly, isolated graphene nanoribbons show photoluminescence in the same wavelength regime.[27] Flat and undisturbed graphene nanoribbons should feature, however, in most cases a zero bandgap.[28] Structural constraints like twisting or bulging of the graphene sheets are likely the inception to an opening of electronic bandgaps.[29]

The non-uniform structure in terms of $sp^2$ or $sp^3$ carbon atoms in the rather large hydrogenated graphene flakes prompts to the question of the origin of its photoluminescence. Recently, the bandgap of highly hydrogenated graphene was spectroscopically determined to be ~4 eV.[30] Direct radiative bandgap deactivation would, therefore, be expected to occur in the ultraviolet range. Photoluminescence is, however, typically observed across the visible and the near infrared regions of the solar spectrum. Therefore, alternative deactivation mechanisms are likely to be operative in the excited state deactivation of hydrogenated graphene.

As discussed in recent studies, "island-type" hydrogenation is likely to govern the hydrogenation of graphene.[5,12,17] When employing, for example, Birch reduction conditions such $sp^2/sp^3$ islands seem to be the favorably formed product.[30] It is postulated that the corresponding $sp^2$ domains form superstructures featuring electronically isolated emissive centers. Still, results stemming from theoretical investigations are contradictive, especially in the context of quantum confinement effects.[7,31] However, it could be shown in *ab initio* quantum mechanical calculations that the electronic properties of π-conjugated domains in hydrogenated graphene are treated similar to those in SWCNTs.[10] In this context, the terminology of SWCNTs can be adapted to hydrogenated graphene (phG) to describe the $sp^2$-domains as (m,n)-strips.

In the context of the aforementioned, we wish to elucidate the origins of photoluminescence in hydrogenated graphene. In particular, the electronic properties of three different luminescent hydrogenated graphene batches were studied by a variety of steady state and time resolved spectroscopic techniques. The results from these spectroscopic assays were used to investigate the stability of phG in charge-transfer assemblies with amphiphilic perylenediimides (PDI).[32] PDIs have been proven to form stable charge transfer complexes with carbon nanotubes, graphene, and carbon nanodots.[33-37] Owing to their extended π-systems, PDIs accept one or two electrons without undergoing significant structural changes. Therefore, PDIs are particularly well suited as an electroactive surfactant to individualize and to stabilize nanostructures that feature aromatic π-systems.

Results and Discussion

Hydrogenation was achieved by means of Birch conditions according to previously reported procedures using different proton sources such as ethanol (**1**), propan-2-ol (**2**), or *tert*-butanol (**3**) and $H_2O$ (**4**) – Scheme S1.[17] In a typical synthesis the pristine spherical graphite was added to a solution of lithium in liquid ammonia (-75°C). After two hours stirring, the proton source was added to the mixture to produce polyhydrogenated graphene (phG) – Scheme S1. To compare the hydrogenation efficiency we applied thermogravimetric analysis coupled with mass spectrometry (TGA-MS). Therefore, the phG samples were heated from rt to 700°C under constant He flow. The phG hydrogenated by water (**4**) exhibited a higher mass loss as well as a two times higher intensity for $H_2$ (m/z 2) than the with alcohols hydrogenated phGs (**1-3**). Thus we conclude that the highest hydrogenation efficiencies were achieved with $H_2O$ (Figures S1 and S2). This result can be traced back to the equilibrium of water and liquid ammonia which facilitates a very slow hydrogenation of the exfoliated graphene sheets and which leads to a different hydrogenation pattern.[17] This different hydrogenation pattern could explain why only for **4** photoluminescence was observed. Furthermore, other differently sized graphites, that is, pyrolitic (**5**) and natural graphite (**6**) were hydrogenated. The hydrogenated pyrolitic graphite (**5**) showed a similar tg profile and hydrogen intensity as the phG **4**. But for phG **6** a decrease in mass loss and hydrogen intensity has been observed. This can be due to the bigger flake size and the worse exfoliation of the graphene sheets. A full characterization of the raw material is given in the Supporting Information.

Transmission electron microscopy was utilized to gather an overview regarding morphology and size distribution. The phGs were dispersed in chloroform and dropcast onto ultrathin carbon films. Throughout the scanned area we found predominantly large accumulations of crumpled carbonaceous sheets with sizes in the µm regime – Figure 1. A large proportion of the nanosized sheet-like structures feature lateral dimensions of several hundred nanometers. Nevertheless, some of the structures were in the order of only a few tens of nanometers.

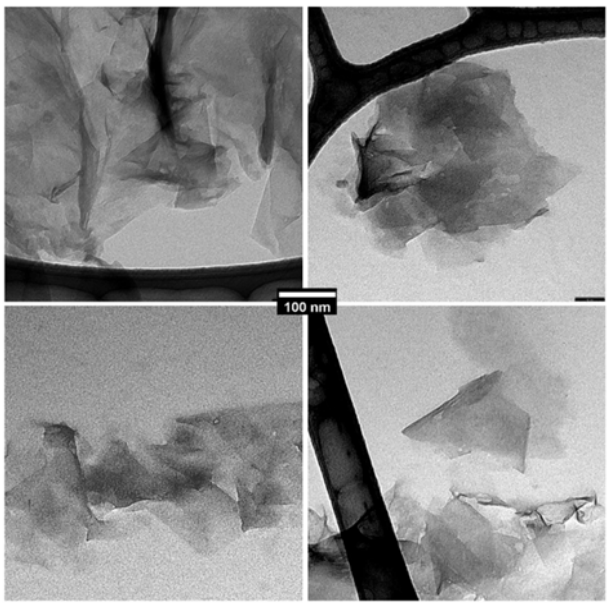

Figure 1: Transmission electron micrographs (80 kV) of **4**. A dispersion of **4** in CHCl$_3$ was dropcast on a Lacey/ultrathin carbon supported TEM grid.

The absorption spectra of **4**, **5**, or **6** either in chloroform or in DMF show weak absorption features at around 4.2 eV – Figures S3 and S4. On the basis of theoretical predictions and experimental results, this is a reasonable value for the first excitonic transition in phG.[30]

In Figure 2, the solid-state 3D photoluminescence spectra of **4**, **5**, and **6** are gathered. All three of them exhibit broad photoluminescence in the region between 450 and 700 nm with maxima at around 540 nm. As derived from the 3D photoluminescence the broad emission is in resonance with excitation in the range from 370 to 500 nm. No particular photoluminescence pattern is, however, appreciable in any of the 3D photoluminescence spectra. Noteworthy is the emission cut off around 355 nm. Using excitation wavelengths below this cut off, namely < 350 nm, fails to produce any appreciable photoluminescence. When comparing the photoluminescence intensity in the three phGs an overall increase in the following order **6** < **5** < **4** is seen. From the spectral and structural resemblance of the spectra we deduce a common denominator for their photoluminescence.

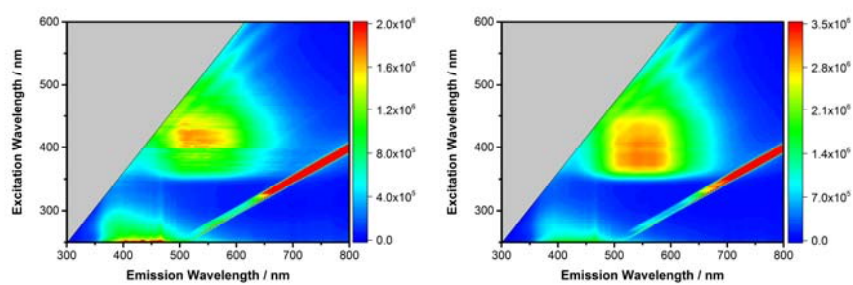

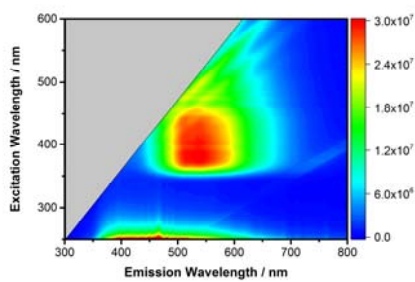

Figure 2: 3D-photoluminescence spectra of **6** (left), **5** (middle), and **4** (right) in the solid state at room temperature.

Correlated excitation-emission spectra taken at several excitation and emission wavelengths in the visible range are collected in Figure 3. As a general trend, excitation within the 370 to 500 nm range generates a rather broad photoluminescence with a maximum at around 540 nm. The photoluminescence intensity is highest upon short wavelength excitation, that is, around 380 nm. Upon excitation at different wavelengths a variation of the photoluminescence, in general, and of its fine structure, in particular, is discernable. Excitation at, for example, 484 nm leads to a photoluminescence with two well defined maxima at 525 and 563 nm. In the corresponding excitation spectrum, a broad spectrum with a distinctive feature at 486 nm evolves.

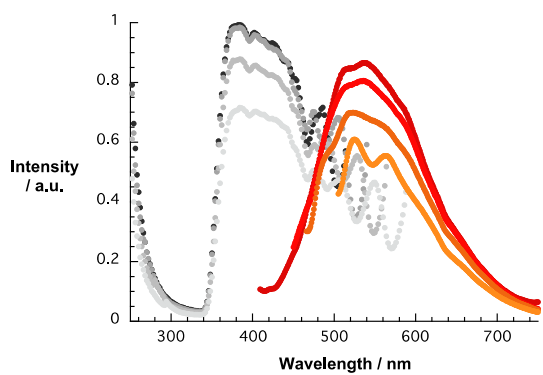

Figure 3: Correlated emission (386 nm (dark red), 426 nm (light red), 450 nm (dark orange), and 484 nm (light orange)) and excitation (525 nm (black), 550 nm (dark grey), 575 nm (intermediate grey), and 600 nm (light grey)) spectra of **4** in the solid state at room temperature.

**Time-Resolved Emission Spectroscopy**

Next, a full-fledged time-resolved emission analysis was conducted starting with photoluminescence lifetime measurements of the three different phGs **4**, **5**, and **6** in the solid state. In line with the steady state photoluminescence spectra, the samples were first excited at 403 nm and the time profiles were taken at 500 and 550 nm – Table 1. At both wavelengths two equally weighted lifetimes, that is, a short and a long-lived component, are found. At 500 nm, the lifetimes are on the order of around 0.4 and 1.7 ns, while at 550 nm, the corresponding values are 0.8 and 3.0 ns. For **4**, **5**, and **6**, resembling lifetimes corroborates once more the common origin of the photoluminescence.

Longer lifetimes, as they were particularly noted in the low energy region, point towards different contributions throughout the spectral window. Therefore, we performed with **4** a time-resolved emission analysis in the wavelength range from 425 to 650 nm – Figure S5 and Table 2. Overall, we note a correlation in terms of relative amplitudes and lifetimes at each wavelength. All of the time profiles were best fit by biexponential decay functions affording a short and a long-lived component. Both components steadily increase from 0.3 and 1.0 ns to 2.0 and 6.6 ns when going from the shorter (~450 nm) to the longer wavelengths (~650 nm).[47] In the next step, we probed the phGs in solution. We focused in our investigations on **4**, since its optical properties resemble that of the other two phGs. Meta-stable dispersions of **4** were prepared by making use of ultrasonication. Different solvents, namely, dimethylformamide (DMF), tetrahydrofuran (THF), toluene, $H_2O$, and chloroform ($CHCl_3$) were tested, but the best results in terms of stability and quantum yields were achieved in DMF and chloroform. Representative absorption spectra of **4** are shown in Figure S6.

**Table 1: Photoluminescence lifetimes of different phGs in the solid state obtained upon 403 nm excitation.**

|   | $\lambda_{PL}$ | $\tau_1$ / ns | $\tau_2$ / ns |
|---|---|---|---|
| **6** | 500 nm | 0.38 (50 %) | 1.7 (50 %) |
|   | 550 nm | 0.75 (40 %) | 2.7 (60 %) |
| **5** | 500 nm | 0.39 (57 %) | 1.7 (43 %) |
|   | 550 nm | 0.85 (56 %) | 3.2 (44 %) |
| **4** | 500 nm | 0.45 (54 %) | 1.8 (46 %) |
|   | 550 nm | 0.93 (56 %) | 3.3 (44 %) |

**Table 2: Photoluminescence lifetimes of 4 in the solid state recorded at different wavelengths upon 403 nm excitation.**

| $\lambda_{PL}$ | $\tau_1$ / ns | | $\tau_2$ / ns | |
|---|---|---|---|---|
| 450 nm | 0.27 | (86%) | 1.05 | (14%) |
| 475 nm | 0.40 | (78%) | 1.59 | (22%) |
| 500 nm | 0.61 | (71%) | 2.20 | (29%) |
| 525 nm | 0.79 | (68%) | 2.68 | (32%) |
| 550 nm | 1.19 | (76%) | 3.85 | (24%) |
| 575 nm | 1.43 | (74%) | 4.40 | (26%) |
| 600 nm | 1.79 | (82%) | 5.75 | (18%) |
| 625 nm | 2.02 | (83%) | 6.60 | (17%) |

At first glance, background absorption, which rises continuously from the low energy (800 nm) to the high energy (300 nm) part, dominates the absorption spectra. A closer look discloses weak features in the range between 380 and 500 nm – Figure S6. The features, for example, for **4** in DMF were also detected in the solid-state excitation spectra – Figure 3. Broad photoluminescent features ranging from 450 to 750 nm with maxima at ~515 nm were detected in THF, chloroform, and DMF but neither in toluene nor in H$_2$O.

The 3D photoluminescence spectra as shown in Figure 4 illustrate the photoluminescence of **4** in DMF dispersions. Like what is seen in the solid state spectra no appreciable fine structure is discernable in the wavelength region from 450 to 750 nm. In contrast to the 3D photoluminescence spectra in the solid state, the emitting states are now resonantly excited at wavelengths < 300 nm. Importantly, < 300 nm matches the energy range of the optical bandgap of phG (4.2 eV). A likely rationale for the different resonant excitation energies is the additional flexibility of phG in dispersions. The latter facilitates vibrational coupling between the electronic states of the sp$^3$ domains and those of the π-conjugated islands. A closer look at the excitation-emission correlation – Figure 5 – enables discerning a clear fine-structure in the excitation spectra.

Next, the photoluminescence lifetimes of **4** were tested in the different dispersions.[48] All samples were excited at 435 nm and the time profiles were recorded at 500 nm. Again, the photoluminescence decays were reasonably fit by biexponential functions yielding lifetimes on the order of 0.4 and 2 ns – Table S2 and Figure S7. The presence of a short 0.3 – 0.5 ns component of large amplitude and a long 1.6 – 2.8 ns component of small amplitude was noted in all solvents. In THF and DMF both components are significantly longer than in toluene and chloroform. As the trends for the lifetimes in the different solvents indicate this effect is attributable to a stabilization based on coordinative interactions between solvent molecules and pHG rather than the solvent polarity.

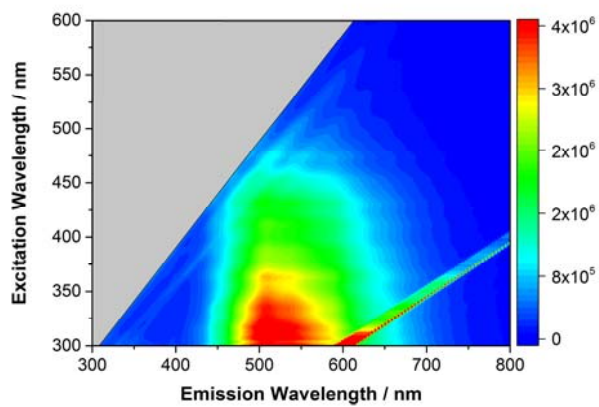

Figure 4: 3D-photoluminescence spectra of **4** dispersed in DMF.

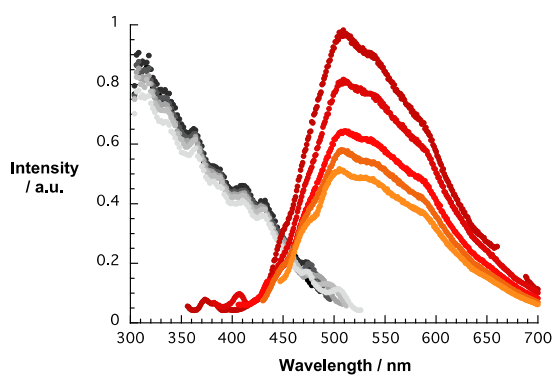

Figure 5: Correlated emission (336 nm (brown red), 362 nm (dark red), 386 nm (light red), 410 nm (dark orange), and 434 nm (light orange)) and excitation (500 nm (black), 510 nm (dark grey), 520 nm (intermediate grey), and 530 nm (light grey)) spectra of **4** dispersed in DMF at room temperature.

To investigate the wavelength dependence we conducted photoluminescence lifetime measurements. The resulting time profiles are shown in Figure S8, from which conclude the same tendency towards longer lifetimes in the longer wavelength regime as seen in the solid-state investigations.

Finally, we turned to femtosecond transient absorption spectroscopy to probe the ultra-fast deactivation mechanisms in pHG. To this end, data were collected for **4** in the solid-state and in DMF dispersions upon 387 nm excitation. Spectra for the solid-state were obtained with samples of **4** dropcasted $CHCl_3$ dispersions onto glass slides and adjusting the optical density to ~0.5. As Figure S10 illustrates, short lived transients in the range between 500 and 700 nm are observed. Immediately after excitation two broad transient minima evolve at around 560 and 650 nm. These transient minima decay rapidly and triphasic with lifetimes of <0.3, ~2.5, and ~100 ps – Table S3 – and independent on the slide. In general, longer lifetimes are detected in the longer wavelength regime.

**Table 3: Photoluminescence lifetimes of 4 dispersed in DMF recorded at different wavelengths upon 435 nm excitation.**

| $\lambda_{PL}$ | $\tau_1$ / ns | | $\tau_2$ / ns | |
|---|---|---|---|---|
| 475 nm | 0.33 | (93%) | 1.63 | (7%) |
| 500 nm | 0.39 | (88%) | 1.79 | (12%) |
| 525 nm | 0.45 | (79%) | 1.88 | (21%) |
| 550 nm | 0.55 | (73%) | 2.27 | (27%) |

As a matter of fact, these lifetimes are in sound agreement with those noted in quantum confined single-walled carbon nanotubes or in graphene.[38] Like in graphene, the observed negative transients is rationalized on the basis of exciton-phonon interactions. Upon laser excitation excitons and phonons are, for example, generated, which form vibrons by interaction through vibrational coupling. The ultra-short component of <0.3 ps, which is, however, shorter than the instrumental response limit is assigned to cooling of hot charge carriers by phonon coupling. The medium lifetime of ~2.5 ps is attributed to the hot-phonon effect.[39] The deactivation of phonons generated upon photon absorption is slowed by coupling with the charge carriers. As a matter of fact, the lifetime determined in our experiments agrees well with phonon deactivations seen in few layer graphene and carbon nanotubes.[40-42] Finally, the long lifetime of ~100 ps is due to the recombination of "relaxed" excitons.

On the basis of the aforementioned results, we hypothesize the existence of geometrically separated islands, which give rise to independent electronic structures as they are seen in SWCNTs and/or graphene nano-ribbons.[43] The well-studied optical properties of SWCNTs provide a good starting point for our discussions. Here, due to the one-dimensional quantum confinement van Hove singularities, that is, discrete transitions between states of electron density, emerge. For SWCNTs, it is well-known that their electronic gap between van Hove singularities scales inversely with their diameters. For example, bandgap transitions for small diameter SWCNTs – (6,4)-SWCNT or (9,1)-SWCNT with diameters 0.68 or 0.75 nm, respectively, are noted in the wavelength regime around 900 nm.[44,45]

Unzipping SWCNTs leads to two-dimensional graphene ribbons. Their width corresponds hereby to the circumference of SWCNTs – Figure S9. A linear plot of the circumference of all known semiconducting SWCNTs versus the band gap transitions is extrapolated in Figure 6 into the visible region. This is, in principle, equivalent to a reversed Kataura plot.[24]

Considering the phG photoluminescence in the 450 to 800 nm range, the corresponding band gap relates to graphene nano-ribbon-like "islands" featuring diameters from 1.1 to 1.75 nm within the phG flakes. Vibronic coupling between the sp³ parts and the sp² islands facilitates manifold electronic deactivation mechanisms in the bulk material. In Figure 6, an illustration of a geometrically relaxed atomic structure of a partially hydrogenated graphene flake with sp² domains in different colors is shown.

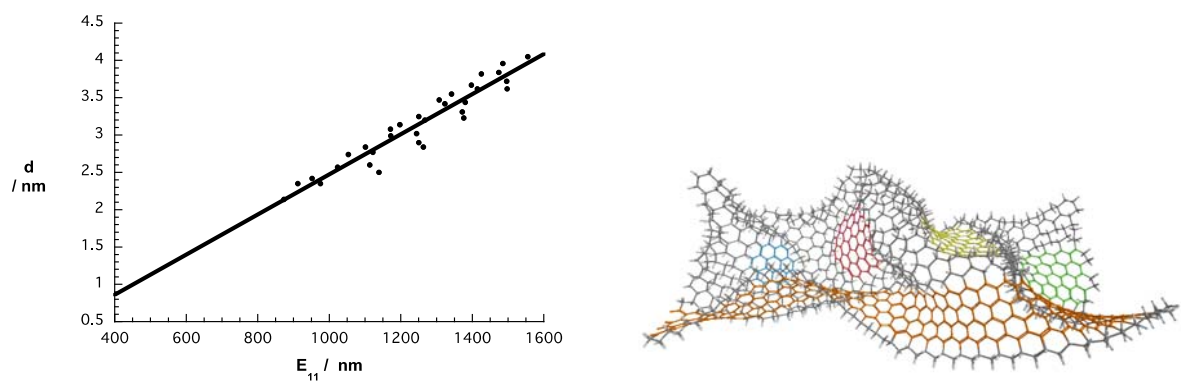

Figure 6: Upper part – a correlation of SWCNT circumferences – equivalent to island width – versus optical band gaps and its extrapolation to small islands in the visible regime. Lower part – illustration of sp²-islands in phG in different colors.

**Dispersion with an electroactive PDI based surfactant**

As a complement to the aforementioned, we tested the dispersability of phG with amphiphilic perylenediimide **PDI$^{6-}$** - Figure S11. The aromatic core of PDIs allows for π-stacking onto π-conjugated carbon nanostructures.[33-37,46] In **PDI$^{6-}$**, two negatively charged Newkome-type dendrons at imido positions provide excellent water solubility.

The formation of the nanohybrids was realized by ultrasonication of sequentially added phG to a $5 \times 10^{-6}$ M solution of **PDI$^{6-}$** in phosphate buffered $H_2O$ and monitored by absorption and emission spectroscopies as shown in Figure 7. Addition of phG to **PDI$^{6-}$** solutions was repeated until no additional changes were observed. The resulting dispersions were subsequently centrifuged at 2, 10, 20 kG.

We note that with increasing amounts of **4** any of the PDI centered absorptions in **PDI$^{6-}$** decrease. At the same time, the background absorption due to phG rises across the spectral range. Important is the intensity reversal in the vibrational fine structure of **PDI$^{6-}$** absorption from an aggregated to a monomeric form without and with **4**, respectively. As such we postulate electronic decoupling between PDIs driven by phG and affording **PDI$^{6-}$/4** hybrids.[4]

After centrifugation the overall absorption decreases to 47%, 37%, and 31% upon centrifugation at 2, 10, and 20 kG, respectively, as a significant amount of the larger phG flakes precipitates during the centrifugation. More important is, however, that the absorption signatures of **PDI$^{6-}$/4** remain unchanged and, thus, underpin the stability of **PDI$^{6-}$/4** hybrids.

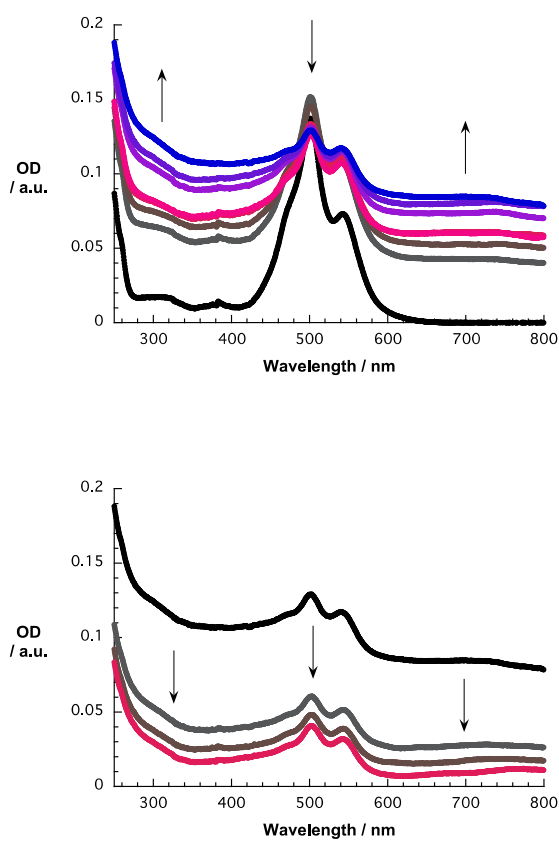

Figure 7: Upper part – absorption spectra of a solution of **PDI$^{6-}$** (black) upon sequential addition of **4** (brown>red>blue) in phosphate buffered H$_2$O. Lower part – absorption spectra of enriched dispersion of **PDI$^{6-}$/4** before (black) and after centrifugation with 2, 10, and 20 kG (brown>red) at room temperature.

The fluorescence of **PDI$^{6-}$** and phG were stimulated selectively at 500 nm and 350 nm, respectively, and monitored during the sequential enrichment as well as after the centrifugation as shown in Figure 8. Here, the PDI centered fluorescence is subject up to a 65% quenching of the original intensity – Figure S12. Moreover, hypsochromic shifts from 550 to 548 nm and from 591 to 589 nm refer to the formation of π-stacks with aligned dipole moments as observed in H-aggregates.[50] When looking at the phG centered photoluminescence a slight increase of the intensity evolves as phG is added to **PDI$^{6-}$** solutions – Figure 8.[51] Throughout the enrichment, the overall intensity of the broad emission feature rises, which is attributable to higher amounts of free phG in the solution. As aforementioned the PDI centered emission is quenched with increasing amounts of phG. This results in a relative decrease of the intensity between 550 and 600 nm – from blue to green in Figure 8. Centrifugation with 2, 10, and 20 kG leads to an overall decrease of the emission intensity along with some qualitative changes. At, for example, higher centrifugation forces a large portion of the ~550-650 nm part diminishes, leaving the ~ 450 nm part as the prominent emission. Upon centrifugation the larger flakes containing larger non-hydrogenated domains are preferably removed from the solution. The smaller flakes with a higher degree of hydrogenation, on the other hand, remain in the solution.

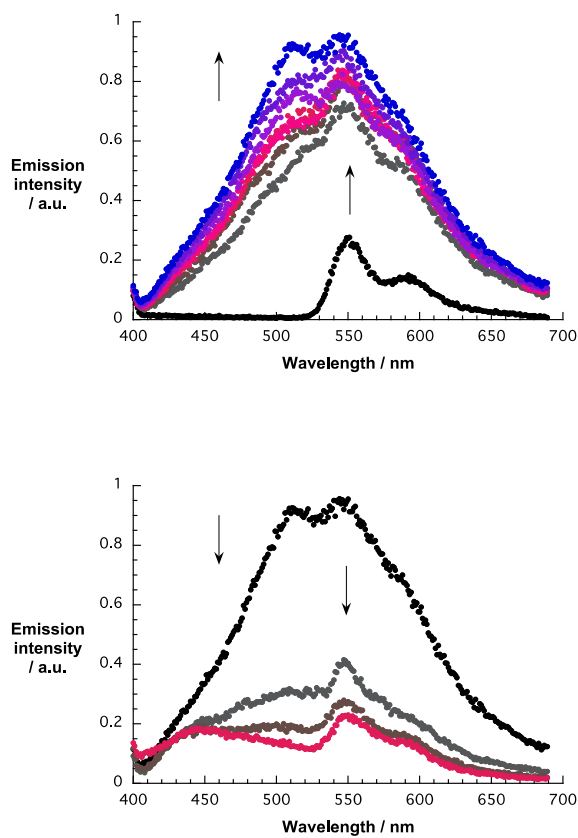

Figure 8: Upper part – photoluminescence spectra of a solution of **PDI**$^{6-}$ (blue) upon sequential addition of **4** (blue>cyan>green) in phosphate buffered H$_2$O. Lower par – photoluminescence spectra of enriched dispersion before (black) and after centrifugation with 2, 10, and 20 kG (brown>red>blue) in phosphate buffered H$_2$O upon excitation at 350 nm at room temperature.

Insights into the nature of the electronic interactions between **PDI**$^{6-}$ and **4** came from fluorescence lifetime measurements. The fluorescence of **PDI**$^{6-}$ was best fit biexponentially with a major lifetime of 4.7 ns and a minor lifetime of 0.3 ns – Figure S13. Similar decay lifetimes were seen in the **PDI**$^{6-}$/**4** hybrid pointing to a static adsorption of **PDI**$^{6-}$ on **4**. Hybrids were obtained *via* the enrichment process described above and emission time profiles were recorded for enriched samples prior and after the centrifugation.

Considering the electron accepting nature of **PDI**$^{6-}$ a unidirectional electron transfer from the π-states of **4** to those of **PDI**$^{6-}$ is implicit. To shed light onto the nature of the electronic communication in **PDI**$^{6-}$/**4** we conducted femtosecond transient absorption spectroscopic measurements with **PDI**$^{6-}$/**4** dispersions. The transient absorption spectra including the corresponding time absorption profiles of **PDI**$^{6-}$ are shown in Figure S14 the supporting information. 530 nm excitation of, for example, **PDI**$^{6-}$ leads to differential absorption changes, which are best characterized as a mix of maxima at 597 nm and minima at 500 nm. The latter relate to bleaching of the ground state absorption. These singlet excited state features decay with 355 ps to recover non-radiatively the ground state. Intersystem crossing is in **PDI$^{6-}$** a minor deactivation channel. **PDI$^{6-}$/4** excitation at 530 nm leads to singlet excited state features similar to those seen in **PDI$^{6-}$**. These transform into two bleaching at 470 and 510 nm and a broad positive transient with a pronounced maximum at 600 nm and several shoulders at 700, 840, and 960 nm – Figure 9. The 470 nm minimum and the 840 and 960 nm maxima at assigned to the one-electron reduced form of **PDI$^{6-}$**.[35] In accordance, we postulate the formation of an intermediate charge transfer product with formation kinetics in terms of charge separation of 3.3 ps and decay kinetics in terms of charge recombination of 62 ps.

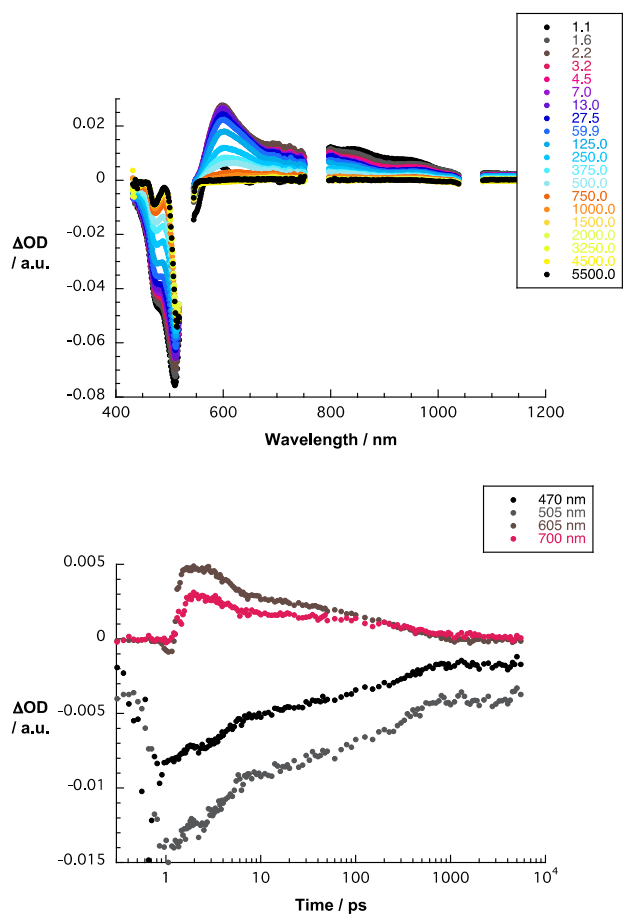

Figure 9: Upper part – differential absorption spectra obtained upon femtosecond pump probe experiments (530 nm) of **PDI$^{6-}$/4** in phosphate buffered H$_2$O with several time delays between 1.1 and 5500 ps at room temperature. Lower part – time absorption profiles of the spectra shown in the upper part at 470, 505, 605, and 700 nm monitoring the excited state decay.

Conclusions

The bandgap in polyhydrogenated graphene (phG) and its photoluminescence renders it useful for applications in optoelectronic devices. Our results, which are based on comprehensive photoluminescence lifetime analyses of phGs in combination with time-correlated single-photon counting spectroscopy, steady-state fluorescence spectroscopy, and femtosecond transient absorption spectroscopy, demonstrate that opening a bandgap in graphene by means of hydrogenation is related to the formation of partially isolated π-conjugated islands. In particular, different starting materials and proton sources for the hydrogenation were used and compared. Photoluminescence is only observed in phG prepared with $H_2O$ as a proton source. All photoluminescent phGs show the same spectral characteristics, although the highest quantum yields were achieved with spherical graphite as starting material.

The broad photoluminescence feature of phGs stretching across the visible part of the spectrum was analyzed in terms of lifetimes and discussed on the basis of quantum confinement. As such, the conclusion drawn from these assays is that graphene islands with diameters in the range from 1.1 to 1.75 nm reveal band gap photoluminescence between 450 and 800 nm. After absorption of a photon (PA) the electrons relax non-radiatively (NRD) *via* vibrational coupling to form excitons at the $sp^2$-islands followed by radiative recombination (Em) of the relaxed excitons on the time scale of ~100 ps – Figure 10.

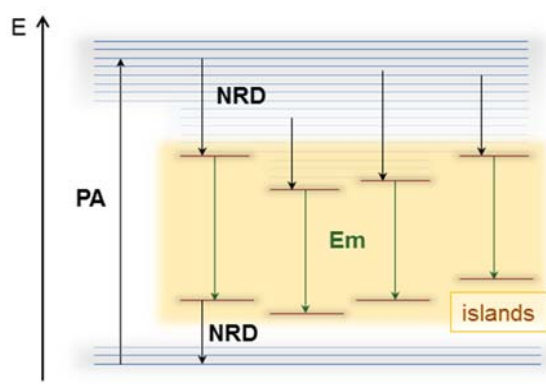

Figure 10: Schematic energy diagram of phG indicating photon absorption (PA), non-radiative decay (NRD), and emissive transitions (Em).

As a complement, phGs were implemented in hybrids with water-soluble electron accepting perylenediimdes (PDI). By virtue of mutual π-stacking and charge transfer interactions with graphene islands, PDIs assisted in stabilizing aqueous dispersion of phG. Implicit in these ground state interactions is the formation of charge separated states once photoexcited. To this end, with the help of femtosecond transient absorption spectroscopy these charge separated states were identified and their lifetimes were determined to be ~300 ps. Polyhydrogenated graphene has great potential as innovative material for applications in optoelectronic devices. For example, establishing a methodology for gaining full control over the hydrogenation of graphene would pave the way to selectively tune its electronic band gap.[11]

**ACKNOWLEDGMENT**

This work was supported by the Deutsche Forschungsgemeinschaft as part of the Excellence Cluster "Engineering of Advanced Materials" and the Collaborative Research Centre SFB 953 "Synthetic Carbon Allotropes". The work leading to these results has received partial funding from the European union Seventh Framework Programme under grant agreement no. 604391 Graphene Flagship. The Bayerische Staatsregierung is also kindly acknowledged for funding granted as part of the "Solar Technologies go Hybrid" initiative. Volker Strauss was supported by the "Universität Bayern e.V."

[47] At 425 nm the lifetimes were below the limits of the instrumental response.
[48] Notable, we were unable to produce stable $H_2O$ dispersions of phG with any appreciable photoluminescence.
[49] The possible decomposition of **PDI**$^{6-}$ upon due to the exposure to ultrasonication was ruled out by reference experiments.
[50] After centrifugation the fluorescence decreases further to 43 % indicating that a significant portion of **PDI**$^{6-}$ is adsorbed to the π-surface of larger phG flakes, which remain in the precipitate.
[51] **PDI**$^{6-}$ shows little absorption at 350 nm, therefore excitation of solely phG is not possible.

# Supporting Information to

# Polyhydrogenated Graphene – Excited State Dynamics in Photo- and Electroactive 2D-Domains

Volker Strauss, Ricarda A. Schäfer, XX, Andreas Hirsch, Dirk M. Guldi

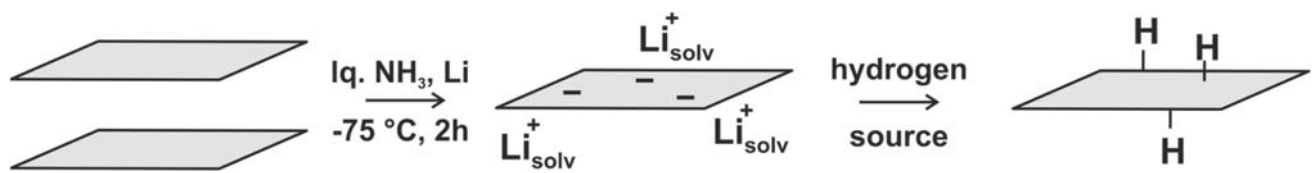

Scheme S1: Synthesis of polyhydrogenated graphene with ethanol (1), propa-2-ol (2), or tert-butanol (3) and water (4) as hydrogen source .

# Thermogravimetric Analysis coupled with Mass Spectrometry

**Table S1: Material parameters**

| material | grade | description | C / % | grain / μm | $d_{bulk}$ / gcm$^{-3}$ | pH | TGA Δm / % |
|---|---|---|---|---|---|---|---|
| 6 | Nat. Pas. | natural flake | --- | 1000 | 0.74 | 7.33 | 2.38 |
| 5 | 2012 | Sri Lankan (natural pyrolitic) | >97 | 70-150 | 0.88 | 6.92 | 4.22 |
| 4 | SGN18 | natural spherical | 99.99 | 20 | 0.92 | 7.12 | 0.54 |

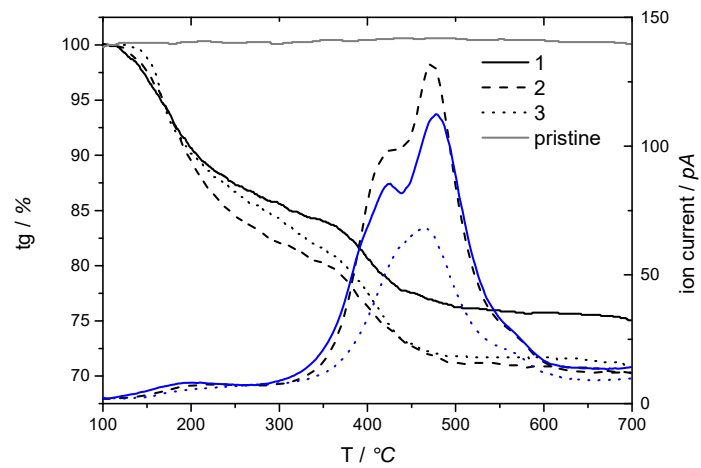

Figure S1: In black: thermogravimetric analysis (10K/min) coupled with mass spectrometry of **1** (solid), **2** (dashed) and **3** (dashed) black under He atmosphere. In blue: mass spectrometric profiles (m/z 2) of **1** (solid), **2** (dashed) and **3** (dashed).

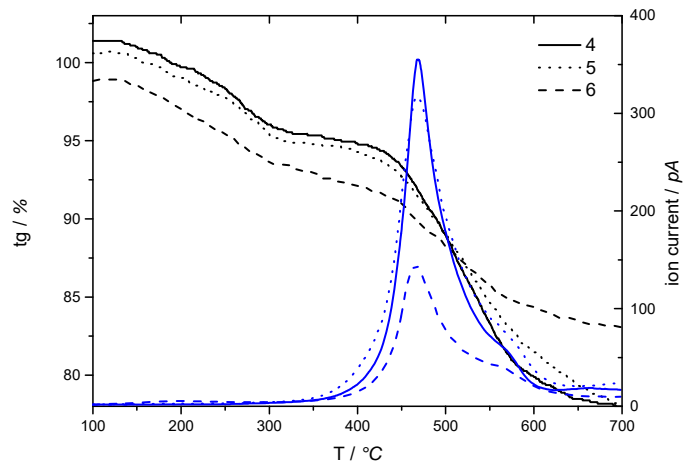

Figure S2: In black: thermogravimetric analysis (10K/min) coupled with mass spectrometry of **4** (solid), **5** (dotted), and **6** (dashed) black under He atmosphere. In blue: mass spectrometric profiles (m/z 2) of **4** (solid), **5** (dotted), and **6** (dashed).

**Steady State Absorption Spectroscopy**

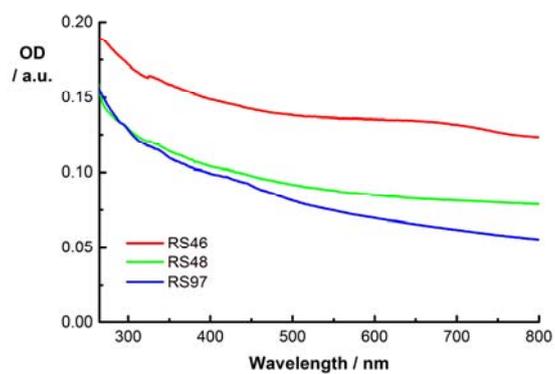

Figure S1: Steady state UV-vis absorption spectra of **4**, **5**, and **6** dispersed in DMF.

# Time-Resolved Emission Spectroscopy

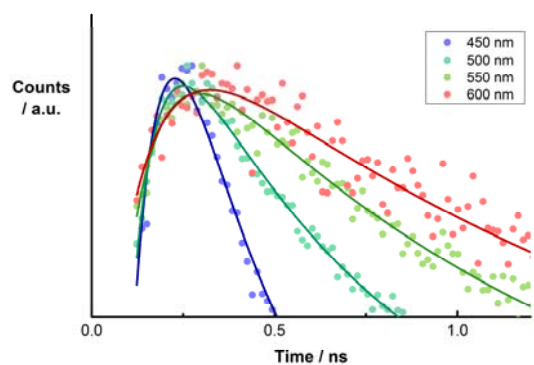

Figure S2: Emission-time profiles of **4** at 450, 500, 550, and 600 nm obtained by time-resolved spectroscopy upon excitation at 403 nm.

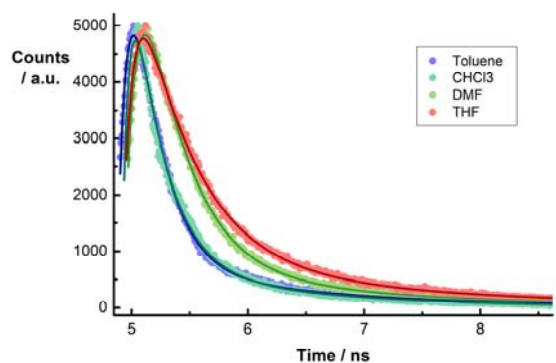

Figure S3: Photoluminescence time profiles of **4** dispersed in different solvents recorded at 500 nm upon 435 nm excitation.

Table S2: Photoluminescence lifetimes of **4** dispersed in different solvents recorded at 500 nm upon 435nm excitation.

|         | $\tau_1$ / ns |       | $\tau_2$ / ns |       |
|---------|---------------|-------|---------------|-------|
| toluene | 0.31          | (90%) | 1.97          | (10%) |
| THF     | 0.55          | (88%) | 2.75          | (12%) |
| CHCl3   | 0.29          | (88%) | 1.56          | (12%) |
| DMF     | 0.47          | (93%) | 2.84          | (7%)  |

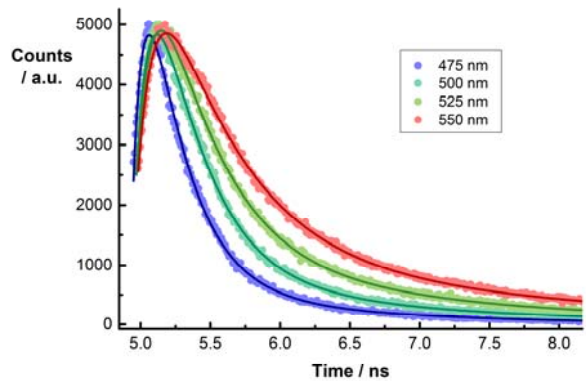

Figure S4: Photoluminescence time profiles of **4** dispersed in DMF recorded at different wavelengths upon 435 nm excitation.

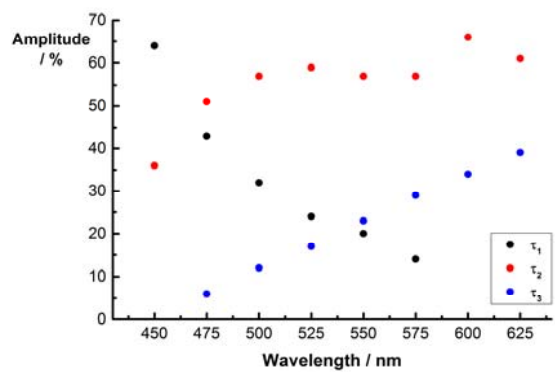

Figure S5: Amplitudes of the lifetimes of **4** obtained by time-resolved spectroscopy upon excitation at 403 nm.

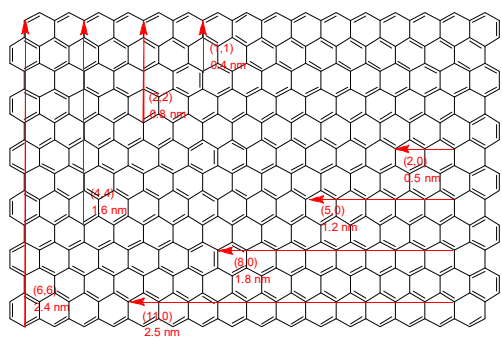

Figure S6: Illustration of a graphene sheet for the determination of the island sizes.

# Femtosecond Transient Absorption Spectroscopy

**Table S3: Lifetimes obtained of 4 by femtosecond transient absorption spectroscopy upon excitation at 387 nm.**

| slide | emission wavelength | $\tau_1$ / ps | $\tau_2$ / ps | $\tau_3$ / ps |
|---|---|---|---|---|
| slide1 | 585 | **<0.3** (14%) | **4.7** (69%) | **120** (17%) |
|  | 650 | **<0.3** (38%) | **5.4** (50%) | **178** (12%) |
| slide2 | 580 | **<0.3** (49%) | **2.8** (35%) | **96** (16%) |
|  | 635 | **<0.3** (60%) | **5.8** (29%) | **194** (11%) |
| slide3 | 550 | **<0.3** (74%) | **1.7** (20%) | **77** (6%) |
|  | 640 | **<0.3** (81%) | **2.2** (14%) | **170** (5%) |
| slide4 | 550 | **<0.3** (61%) | **1.9** (30%) | **78** (9%) |
|  | 580 | **<0.3** (77%) | **1.7** (18%) | **92** (5%) |
|  | 635 | **<0.3** (73%) | **2.0** (20%) | **101** (8%) |
| slide5 | 550 | **<0.3** (60%) | **2.3** (29%) | **69** (11%) |
|  | 650 | **<0.3** (66%) | **1.9** (25%) | **98** (9%) |

## Dispersion with an electroactive PDI based surfactant

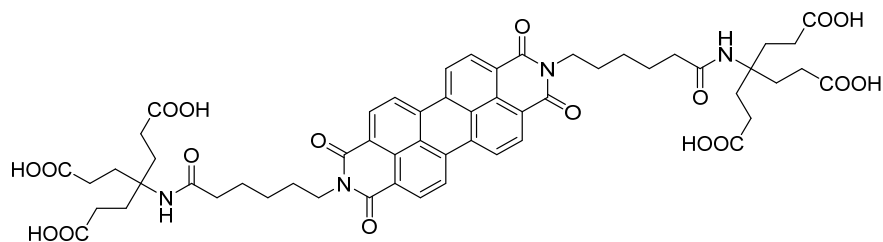

Figure S7: Molecular structure of **PDI**[6-].

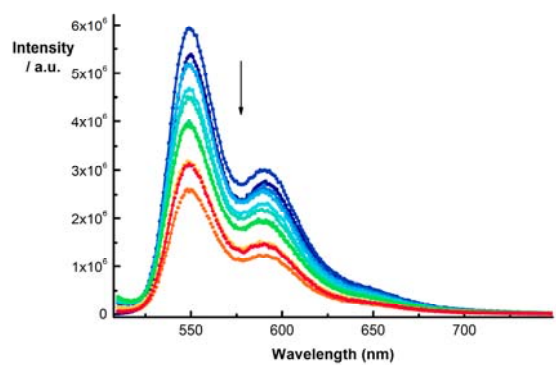

Figure S8: Steady state emission spectra of **PDI**[6-] (blue) upon sequential addition of **4** (blue>cyan>green) and after centrifugation with 2, 10, 20 kG (orange>red) in phosphate buffered $H_2O$ upon excitation at 500 nm at room temperature.

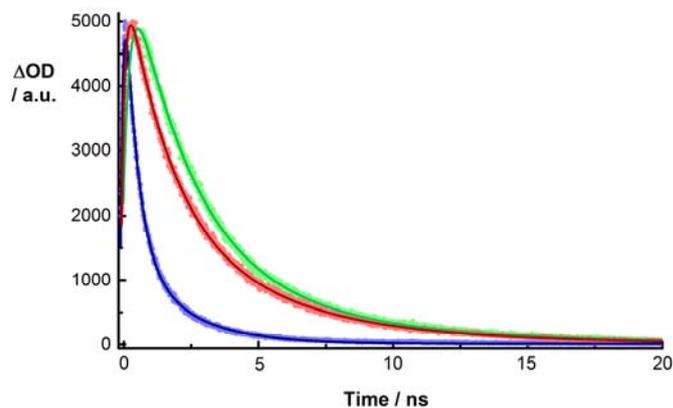

Figure S9: Photoluminescence time profiles of **PDI**[6-] (green), **PDI**[6-]/**4** (red), and 4 (blue) in phosphate buffered $H_2O$ recorded at 550 nm upon 500 nm excitation.

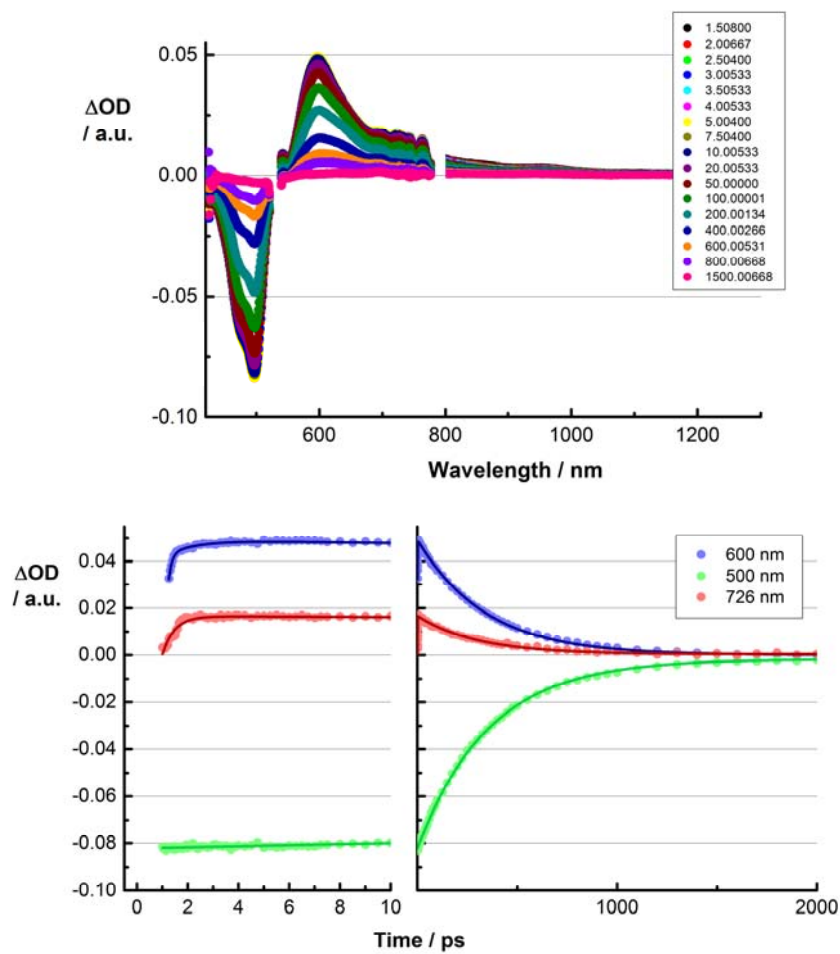

Figure S10: Upper part – differential absorption spectra obtained upon femtosecond pump probe experiments (530 nm) of **PDI**[6-] in phosphate buffered $H_2O$ with several time delays between 0 and 5000 ps at room temperature. Lower part – time absorption profiles of the spectra shown in the upper part at 500 nm (green), 600 nm (blue), and 725 nm (red) monitoring the excited state decay.